\documentclass[10pt,aps,pra]{revtex4}

\usepackage{amsmath}
\usepackage{graphicx}

\begin{document}

\title{Homogeneous solutions for elliptically polarized light in a cavity
containing materials with electric and magnetic nonlinearities}
\author{Daniel A. M\'{a}rtin and Miguel Hoyuelos}

\affiliation{Departamento de F\'{\i}sica, Facultad de Ciencias Exactas y
Naturales, Universidad Nacional de Mar del Plata and Instituto de
Investigaciones F\'{\i}sicas de Mar del Plata (Consejo Nacional de
Investigaciones Cient\'{\i}ficas y T\'{e}cnicas), Funes 3350, 7600 Mar del
Plata, Argentina}

\begin{abstract}
We study evolution equations and stationary homogeneous solutions for
electric and magnetic field amplitudes in a ring cavity with flat
mirrors. The cavity is filled with a  positive or negative refraction
index material with third order Kerr-like electric nonlinearities and
also magnetic nonlinearities, which can be relevant in metamaterials.
We consider the degree of freedom of polarization in the incident
beam. It is found that considering a magnetic nonlinearity increases
the variety of possible qualitatively different solutions. A
classification of solutions is proposed in terms of the number of
bifurcations.  The analysis can be useful for the implementation of
optical switching or memory storage using ring cavities with non
linear materials.
\end{abstract}

\maketitle


\section{Introduction}

During the last decade, progress in the development of composite
materials allowed the experimental observation of new optical
properties, such as negative refraction index \cite{smith,shelby}.
These materials have both negative dielectric permittivity and
negative magnetic permeability, a property not found in natural
materials. Materials with negative refraction index (NRM) have many
interesting new features, and a great number of applications have
been proposed based on their novel properties.

 On the other hand, ring cavities filled with nonlinear media have
been proposed for information storage and also as the optical analog
of electrical transistors \cite{moloney}. Conditions for modifying and
tuning optical bistability in these cavities are being researched
\cite{Kim,anton}. It is known that electromagnetic wave propagation
in a Kerr-type nonlinear material with a positive refraction index
(PRM) can be described by an order parameter equation  of the
nonlinear Schr\"{o}dinger type and the same type of equation  can be
extended to NRM \cite{kockaert,kockaert2,wen,scalora,lazarides}.

A recent work \cite{Merlin} shows that arguments previously used to
neglect the magnetic response in a PRM do not hold for NRM, where
both the polarization and the magnetization may take non-negligible
values. Even, it has been shown \cite{zharov,lapine} that a composite
metamaterial with negative refraction index could develop a nonlinear
macroscopic magnetic response. This means that although the medium
has a negligible magnetic nonlinearity, periodic inclusions produce a
nonlinear effective response when the wavelength is much larger than
the periodicity of the inclusions.

In this paper we analyze the equations describing the electric  and
magnetic fields in a ring cavity with flat mirrors that contains a
NRM or a PRM with electric and magnetic nonlinearities. In a previous
work \cite{nosotros}, we limited ourselves to a linearly polarized
incident beam, now we focus on analyzing what happens when the light
polarization is taken into account, making it necessary the double of
equations to describe the system. The main objective of this paper is
to explore the various types of stationary homogeneous solutions that the system can display, depending on the parameters that describe
it. We also perform a brief analysis of the behavior of linearly
unstable solutions.

The paper is organized as follows. In section \ref{Equ} we present
the evolution equations for the amplitude of the electric and
magnetic fields and the reduction to two coupled Lugiato Lefever
equations for a ring cavity. In section \ref{SHE} we analyze
homogeneous stationary solutions. The type and number of bifurcations
that can occur, depending on the parameters, are also analyzed.
Finally, in section \ref{simul}, numerical integration is performed
in order to analyze the temporal evolution of the unstable solutions.
In section \ref{concl}  we present our conclusions.

\section{Cavity Equations}
\label{Equ}

The equations often used for describing the behavior of electric and
magnetic fields, at the plane perpendicular to propagation of light
in a cavity, are of the type of Lugiato-Lefever equation (LL, see
\cite{lugiato}) with generalized parameters. LL equation is a simple
mean field model, and has been useful for the analysis of pattern
formation in cavities containing a Kerr medium that is driven by a
coherent plane wave \cite{hoyuelos,firth}.

First, we analyze the problem of free propagation (without mirrors)
of an electromagnetic wave in a nonlinear material, either PRM or
NRM, and then we use the resulting equations to obtain the behavior
in the cavity.

\subsection{Evolution Equations in the material}
    \label{equ1}

We consider a plane wave of arbitrary polarization and frequency $
\omega_0 $. We assume that the electric and magnetic fields are in
the plane $x$-$y$ and the wave propagates in the $z$ axis. The
starting point is Maxwell's equations and constitutive relations for
the electric displacement, $ \mathbf{D} = \epsilon_0 \mathbf{E} +
\mathbf{P} $, and magnetic induction, $ \mathbf{B} = \mu_0 \mathbf{H}
+ \mu_0 \mathbf{M} $. An isotropic metamaterial is analyzed with the
third-order Kerr nonlinear response. The nonlinear relationship
between the polarization of the material and the electric field is
\cite{moloney}:
\begin{eqnarray}
P_i(t) &=& \epsilon_0 \int_{-\infty}^\infty
\chi^{(1)}_{E,ij}(t-\tau)\, E_j(\tau)\, d\tau \nonumber \\
&& + \epsilon_0 \int_{-\infty}^\infty
\chi^{(3)}_{E,ijkl}(t-\tau_1,t-\tau_2,t-\tau_3)\, E_j(\tau_1)
E_k(\tau_2) E_l(\tau_3) \, d\tau_1 d\tau_2 d\tau_3, \label{pol}
\end{eqnarray}
where $i,j,k$ and $l$ correspond to Cartesian axes, and take values
$x$, $y$ or $z$; repeated indexes involve addition.
$\chi^{(1)}_{E,ij}$ is a rank 2 tensor describing linear electric
behavior in the material and $\chi^{(3)}_{E,ijkl}$ accounts for third
order nonlinearities.

Zharov \emph{et al} \cite{zharov} analyzed properties of a
microstructured metamaterial made of metallic wires and split ring
resonators (SRR) embedded into a Kerr permittivity material. They
found out that, in such a material, the electric fields around the
SRR could be very strong. In that case, the electric nonlinearity
produces an effective magnetic nonlinearity $\mu_{eff}$. Expanding
$\mu_{eff}$ in powers of the incident magnetic field, to the lowest
non constant order, a Kerr-like magnetization can be found. Thus, a
similar relationship to (\ref{pol}) is proposed for magnetization and
magnetic field (with similar definitions for $\chi^{(1)}_{M}$ and
$\chi^{(3)}_{M}$):
\begin{eqnarray}
M_i(t) &=& \int_{-\infty}^\infty \chi^{(1)}_{M,ij}(t-\tau)\,
H_j(\tau)\, d\tau \nonumber \\
&& +  \int_{-\infty}^\infty
\chi^{(3)}_{M,ijkl}(t-\tau_1,t-\tau_2,t-\tau_3)\, H_j(\tau_1)
H_k(\tau_2) H_l(\tau_3) \, d\tau_1 d\tau_2 d\tau_3. \label{pol2}
\end{eqnarray}
Many proposed metamaterials consist of the repetition of a cubic unit
lattice which has the same form when it is looked from any of its
faces (for example, materials using spherical inclusions
\cite{Holloway}, spherical inclusions and wires \cite{Cai} but also
cells specifically designed to be symmetric \cite{Koschny}), making
optical properties to be the same in any two orthogonal directions.
If the wavelength of incident light is much larger than the cell
length, the material is expected to be isotropic. So, we assume
that the material is  cubic centrosymmetric and isotropic. As the
components of the tensors $\chi^{(3)}_{E/M}$ take part only on field
convolutions, the tensors are defined, without loss of generality, to
be symmetrical for a large number of indexes and argument
permutations, similar to what has been done in \cite{moloney}, Ch.\
2, Sect.\ 2d.

We apply the classical multiple scales perturbation technique, and we
assume that light is quasi-monochromatic and can be represented by a
plane wave, with frequency $\omega_0$ and wavenumber $k_0$. It
travels thought  the $z$ axis, and  is modulated by a slowly varying
amplitude. The amplitude depends on position $\mathbf{R} = (X, Y, Z)$
and time $T$, variables that have characteristic scales much larger
than the scales given by $1/k_0 $ and $ 1 / \omega_0 $. In summary,
the fields $ \mathbf{E} $ and $ \mathbf{H} $ can written as
\begin{eqnarray}
\mathbf{E} &=& \boldsymbol{\mathcal{E}}(\mathbf{R},T) e^{i(k_0 z -
\omega_0 t)} + c.c. \nonumber \\
\mathbf{H} &=& \boldsymbol{\mathcal{H}}(\mathbf{R},T) e^{i(k_0 z -
\omega_0 t)} + c.c. \label{ansatz}
\end{eqnarray}
where $ \boldsymbol{\mathcal{E}} $ and $ \boldsymbol{\mathcal{H}} $
are slowly varying amplitudes. Similar relationships are defined for
$P$ and $M$; the corresponding amplitudes can be written as (for the
polarization of material see, for example, \cite{geddes})
\begin{eqnarray}
\boldsymbol{\mathcal{P}} &=& 3 \chi^{(3)}_E \left[A_E
(|\mathcal{E}_x|^2+|\mathcal{E}_y|^2) \boldsymbol{\mathcal{E}}+ B_E/2
(\mathcal{E}_y^2 + \mathcal{E}_x^2) \bar{\boldsymbol{\mathcal{E}}}
\right] \nonumber  \\
\boldsymbol{\mathcal{M}} &=& 3 \chi^{(3) }_M \left[A_M
(|\mathcal{H}_x|^2+|\mathcal{H}_y|^2) \boldsymbol{\mathcal{H}}+ B_M/2
(\mathcal{H}_y^2 + \mathcal{H}_x^2)
\bar{\boldsymbol{\mathcal{H}}}\right],
\end{eqnarray}
where the bar stands for complex conjugate, $\chi^{(3)}_{E/M}=
\chi^{(3)}_{E/M,xxxx}$ and parameters $A_{E/M}$ and $B_{E/M}$ are
given by
\begin{eqnarray}
A_{E/M} &=&
\frac{\chi^{(3)}_{E/M,yyxx}+\chi^{(3)}_{E/M,yxyx}}{\chi^{(3)}_{E/M}}
\nonumber \\
B_{E/M} &=& 2 \frac{\chi^{(3)}_{E/M,yxxy}}{\chi^{(3)}_{E/M}}.
\end{eqnarray}
In the previous expressions we are using the notation
$\chi^{(3)}_{E/M,abcd}$ (without explicit dependence) to identify the
Fourier transform of $\chi^{(3)}_{E/M,abcd}(t_1,t_2,t_3)$ evaluated
in $(\omega_0,\omega_0,-\omega_0)$.

We define the wavenumber $k(\omega) = \omega n(\omega) / c $; $k'$
and $k''$ are the  derivatives of $k(\omega)$ at $\omega=\omega_0$;
the refraction index is $ n(\omega) = \pm \sqrt {\epsilon_r (\omega)
\, \mu_r (\omega)} $ (which takes a negative value when both $
\epsilon_r $ and $\mu_r  $  are negative \cite{veselago}), where
$\epsilon_r (\omega) = 1 + \chi^{(1)}_E (\omega)$ and $\mu_r(\omega)
= 1 + \chi^{(1)}_M(\omega)$ are the relative permittivity and
relative permeability, respectively.

The details of the implementation of multiple scales technique in our
case, are essentially the same as those described in reference
\cite{moloney}, Sect.\ 2k. After this process, we obtain nonlinear
Schr\"{o}dinger equations for the envelopes of electric and magnetic
fields defined in the plane perpendicular to the $z$ axis. Applying a
change of variables to use circularly polarized components,
$\sqrt{2}\mathcal{E}_\pm=\mathcal{E}_x \pm i \mathcal{E}_y$,
$\sqrt{2}\mathcal{H}_\pm=\mathcal{H}_y \mp i \mathcal{H}_x$, and a
coordinate transformation given by $\xi=z$, $\tau = t-k'z$, the
following equations of evolution in the material can be found

\begin{eqnarray}\label{a1}
{\partial \mathcal{E}_\pm \over
\partial \xi} &=& - k''(\omega_0){i \over 2}{\partial^{2} \mathcal{E}_\pm
\over \partial t^{2}} + {i \over 2k_0} \nabla_\perp^{2}
\mathcal{E}_\pm  + {3 k_0 i \over 2}\left[ {\chi^{(3)}_E \over
\varepsilon_r} (A_E
|\mathcal{E}_\pm|^2+(A_E+B_E)|\mathcal{E}_\mp|^2)
\right. \nonumber \\
&& \left. + {\chi^{(3)}_M \over
\mu_r}(A_M|\mathcal{H}_\pm|^2+(A_M+B_M)|\mathcal{H}_\mp|^2)\right]
\mathcal{E}_\pm
\end{eqnarray}

\begin{eqnarray}\label{a2}
{\partial \mathcal{H}_\pm \over
\partial \xi} &=& - k''(\omega_0){i \over 2}{\partial^{2} \mathcal{H}_\pm
\over \partial t^{2}}+{i \over 2k_0}  \nabla_\perp^{2}
\mathcal{H}_\pm + {3 k_0 i \over 2}\left[ {\chi^{(3) }_E \over
\varepsilon_r} (A_E |\mathcal{E}_\pm|^2+(A_E+B_E)|\mathcal{E}_\mp|^2)
\right. \nonumber \\
&& \left. + {\chi^{(3)}_M \over
\mu_r}(A_M|\mathcal{H}_\pm|^2+(A_M+B_M)|\mathcal{H}_\mp|^2)\right]
\mathcal{H}_\pm
\end{eqnarray}
where the approximation $k_0^2\simeq \omega_0^2 \mu_0 \mu_r
\varepsilon_0 \varepsilon_r $, valid for low dispersion fields, is
used.  For simplicity of notation we  write $\mu_r(\omega_0)=\mu_r$ and
$\varepsilon_r(\omega_0)=\varepsilon_r$.

Linear and nonlinear dissipation is neglected, so that
$\chi^{(1)}_{E/M} $ and $\chi^{(3)}_{E/M} $ are real valued
quantities. This is an often used approach in conventional optics,
but dissipation could play an important role in metamaterials.
Nevertheless, as we comment in the next section, there are ways to
reduce dissipation in metamaterials even to negligible levels.  On
the other hand, a generalization of the model for small dissipation is
also possible \cite{nosotros}.

\subsection{Cavity effects}
    \label{equ2}

\begin{figure}
\includegraphics{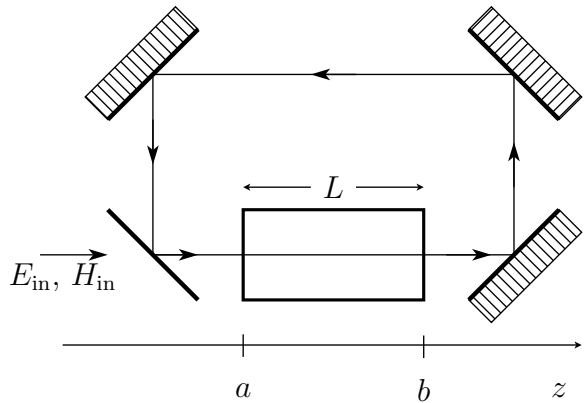}
\caption{Scheme of the ring cavity.} \label{cav}
\end{figure}

We consider a ring cavity with plane mirrors and with a nonlinear
material as shown in Figure \ref{cav}.  At the input mirror, the
incident fields are $E_\mathrm{in}$ and $H_\mathrm{in}$, and the
reflection and transmission coefficient are $ r_i $ and $ t_i $
respectively; $T_{r}$ is the round trip time  and $L$ is the
nonlinear material length. The detuning between the incident field
and the  cavity mode is $\theta = \phi \mod 2\pi$ where $\phi$ is the
phase accumulated by the wave around the cavity, including the
effects of reflections in the mirrors. We study the cavity near
resonance, so the mismatch is small and $e^{i \theta} \simeq 1 + i
\theta$.

From Maxwell equations, assuming that nonlinearities are small, it is
obtained that, in a given medium (different from vacuum), the
relationship between electric and magnetic fields is approximately
$\mathcal{H}_\pm= \eta \mathcal{E}_\pm$ while in vacuum the relation
is $\mathcal{H}_\pm= \eta_0 \mathcal{E}_\pm$, where $\eta$ and
$\eta_0$ are the inverses of the material and vacuum impedances,
respectively. If we consider media with similar impedances
($\eta \simeq \eta_0$), the wave is mainly transmitted, generating a
negligible reflected wave .  The transmitivity coefficients at both
ends of the material are $t_\pm= 1 \pm {\eta_m-\eta_0 \over
\eta_m+\eta_0 }$, where $t_+$ holds for $\mathcal{E}$ at point $b$
(see Fig.\ \ref{cav}) and for $\mathcal{H}$ at point $a$, and $t_-$
holds for $\mathcal{H}$ at $b$ and for $\mathcal{E}$ at $a$.

We can obtain an equation of evolution for electric and magnetic
fields in the cavity and it  can be shown that under the above
conditions, the proportionality between them is maintained, so the
system is well described knowing only the electric field. The
procedure is analogous to that performed in \cite{nosotros}. It is
important to mention that from now on we are looking at cases where
transverse spatial dependence can be neglected, i.e.
$\nabla_\perp^{2} \mathcal{E}_\pm \simeq 0$. We get
\begin{eqnarray}\label{aa14}
 T_{r}{\partial \mathcal{E}_\pm \over \partial t} &=& t_i E_{in \pm}
 + (\rho-1 +\rho i \theta ) \mathcal{E}_\pm  +{3 \rho i L k_0\over 2
 \varepsilon_r} \left[ (  t_+^2  \eta^2 \chi^{(3)}_ M  B_M + t_-^2
 \chi^{(3)}_E  B_E  )|\mathcal{E}_\mp|^2 \right. \nonumber \\ &&
 \left. + ( t_+^2 \eta^2 \chi^{(3)}_M A_M + t_-^2 \chi^{(3)}_E A_E
 )(|\mathcal{E}_+|^2+|\mathcal{E}_-|^2) \right] \mathcal{E}_\pm.
\end{eqnarray}
In order to reduce the number of parameters, an equation with
adimensional quantities can be obtained using the following change of
variables
\begin{eqnarray}
t' = {t (1-\rho) \over T_r } && \Theta = {-\rho \theta\over (1- \rho)
} \label{ch1} \\
A_\pm = {\mathcal{E}_\pm\over \mathcal{E}_{t} }
\sqrt{L k_0 \rho \over 2  \varepsilon_r(1- \rho)}  && A_{in\pm} =
{t_i E_{in\pm} \over \mathcal{E}_{t} (1-\rho)} \sqrt{L k_0 \rho \over
2 \varepsilon_r (1- \rho)}\label{ch2}
\end{eqnarray}
where $\rho = r_i t_+ t_- \lesssim 1$ so that $1-\rho$ is a small
number of the order of $\theta$, and $\mathcal{E}_{t}$ is a
characteristic field given by $1/\sqrt{|3 \chi^{(3) }_M t_+^2 \eta^2
+3 \chi^{(3) }_E   t_-^2 |}$. \
It is useful to define the parameter
$\bar{B}$ as
\begin{equation}
\bar{B}= { \eta^2 \chi^{(3)}_ M  t_+^2 B_M + \chi^{(3) }_E t_-^2 B_E
\over  \eta^2 \chi^{(3)}_M  t_+^2  + \chi^{(3) }_E   t_-^2 }
\end{equation}
that can be understood as an average of the electric and magnetic
constants $B_E$ and $B_M$. In the general case, using symmetry
arguments, it can be shown that $A_E+B_E/2=A_M+B_M/2=1$. In Ref.\
\cite{boyd} (see p.\ 197), three possible cases are mentioned for
only electric nonlinearities (i.e. $A_M=B_M=0$): $B_E=0$
($\bar{B}=0$), for electrostriction, that does not act at optical
frequencies; $B_E=6 A_E$ ($\bar{B}=3/2$) for molecular orientation
effects; and $B_E=A_E$ ($\bar{B}=2/3$) in materials with electronic
response far from resonance frequency.  In the rest of our work, we
will find analytical results for any value of $\bar{B}$, but we will
limit our numerical results to the case $0 \leq \bar{B} \leq 2 $,
which corresponds to situations in which $A_{E/M} \geq 0$,  and where
$\chi_E^{(3)}$ has the same sign as $\chi_E^{(3)}$. The motivation of
the last choice is a result of Ref.\ \cite{zharov}, where the authors
derive an effective Kerr-like nonlinear magnetization for a split
ring resonator; it can be written as
$\mu_{\mathrm{eff}}=\mu_{\mathrm{eff}0}+K H^2$, where $K$ is a
complex number related to geometrical factors, which tends to a real
quantity with the same sign as the electric nonlinearity, when
conductivity tends to $\infty$. Eq.\ (\ref{aa14}) finally reads:
 \begin{eqnarray}\label{evol}
{\partial A_\pm  \over \partial t'}= A_{in\pm} -(1+i \Theta) A_\pm +
i \alpha  (|A_\pm|^2
 (1-\bar{B}/2)+ |A_\mp|^2 (1+\bar{B}/2)) A_\pm,
\end{eqnarray}
where $\alpha$ is the sign of $\chi^{(3) }_M  t_+^2 \eta^2 +
\chi^{(3)}_ E t_-^2 $. Let us note that dispersion effects,
proportional to $k''$, are not included, neither in Eq.\ (\ref{aa14}) nor in Eq.\ (\ref{evol}). The reason is that
dispersion generates a term in (\ref{evol}) proportional to the small
parameter $1-\rho$ used in the scaling (\ref{ch1}) and (\ref{ch2}),
so that it can be neglected with respect to the other terms.

Considering transverse spatial effects would produce the inclusion of
a Laplacian in Eqs.\ (\ref{evol}), transforming them into Lugiato
Lefever equations. Complex conjugate fields describe the same
physical system and the corresponding equation is equal to
(\ref{evol}) with $\Theta$ and $\alpha$ with opposite signs,
therefore the relevant parameter is $\alpha \Theta$. Considering a
metamaterial in this analysis justifies the inclusion of the magnetic
nonlinearity, that allows a wider range of possible values of
$\bar{B}$.  The sign of the refraction index does not appear
explicitly in (\ref{evol}), but it would appear in a diffraction term
if we would take into account transverse spatial effects
\cite{nosotros}.

Up to now, losses have not been taken into account. In the Gigahertz
range, it is possible to build an NRM with small (and even
negligible) imaginary parts of $\epsilon_r$ and $\mu_r$  as was shown
in, for example, \cite{liu,wang}, but for higher frequencies the
former is not true. Many proposals for reducing losses at high
frequencies are being researched, such as
\cite{dolling,shalaev,popov}. If dissipation is not strong, Eq.\
(\ref{evol}) can be generalized in a similar way to what has been
done in a previous work \cite{nosotros}. Taking $k(\omega_0) = k_0 +
i k_I$, if $k_I \ll k_0$, the term $-k_I \mathcal{E_\pm}$ has to be
added to the right hand side of equation (\ref{a1}),  $-k_I
\mathcal{H_\pm}$ has to be added to the right hand side of equation
(\ref{a2}) and $-L k_I \mathcal{E_\pm}$ has to be added to the right
hand side of equation (\ref{aa14}). An adequate change of variables
is easily found replacing $(1- \rho)$ by $(1- \rho+ \rho L k_I)$ in
(\ref{ch1}) and (\ref{ch2}), so that (\ref{evol}) is still valid.

\section{Homogeneous Stationary Solutions} \label{SHE}

Looking for stationary homogeneous solutions in (\ref{evol}), we
obtain
\begin{eqnarray}
\label{estac} A_{in\pm} = [1+ i \Theta- i \alpha
(1-\bar{B}/2)|A_\pm|^2-i \alpha (1+\bar{B}/2) |A_\mp|^2 ]A_\pm.
\end{eqnarray}
From this equation, using that $\alpha^2=1$ we can write
\begin{eqnarray}\label{IntensETA}
 \lambda I_{in} &=& \left(1+ [\alpha \Theta-   (1-\bar{B}/2) I_+ -
 (1+\bar{B}/2)I_-]^2 \right)I_+ \nonumber \\
 (1-\lambda) I_{in} &=& \left(1+ [\alpha \Theta- (1-\bar{B}/2)I_- -
 (1+\bar{B}/2) I_+]^2\right)I_-
\end{eqnarray}
where $I_\pm= |A_\pm|^2$, $I_{in}= |A_{in+}|^2 +|A_{in-}|^2 $, and
$\lambda= |A_{in+}|^2/I_{in}$ is a quantity that measures the
polarization of the incident beam, related to the ellipticity $\chi$
by $\lambda= \cos^2(\chi/2)$.

Given a set of parameters ($\alpha \Theta, \bar{B}, \lambda$),
solution sets $(I_{in}, I_+,I_-)$ can be obtained. By definition,
$0\leq \lambda \leq 1$, but, for each solution with $\lambda > 1/2$
there is another solution that can be found  exchanging
$\mathcal{E}_+$ with $\mathcal{E}_-$, so, we work with $1/2 \leq
\lambda \leq 1$.

It was found that the type of possible solutions varies according to
the parameters of the system. We present a list of them for linear,
circular and elliptic polarization.

\subsection{Linear polarization, $\lambda = 1/2$}

\begin{figure}
\includegraphics[width=12cm]{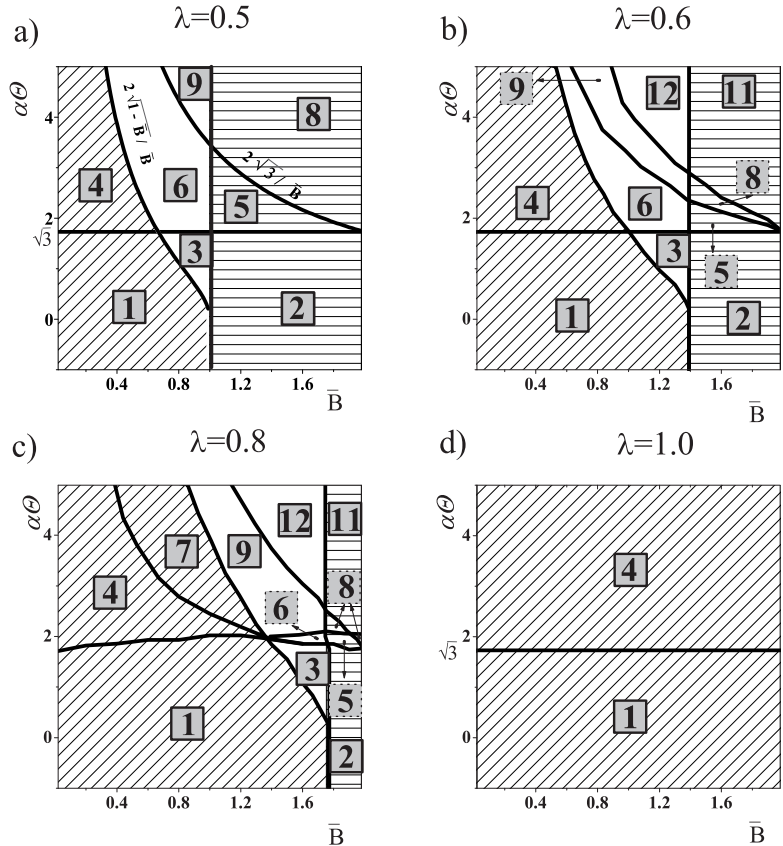} \caption{
Types of solutions in phase space $ \alpha \Theta - \bar{B} $. For
$\lambda = 1/2$ (a) and $\lambda = 1$ (d), the number in each region
is given by $N = 1 + n_p +3 n_s$, where $n_p$ is the number of
pitchfork bifurcations and $n_s$ is the number of bistability
regions given by a pair of saddle node bifurcations. The analytic
expressions for the limits between regions are shown in some cases.
For elliptical polarization (b) and (c), we show a numerical
exploration of possible solutions, and we use the number $ N = 1 +
n_p'+3 n_s $ to identify regions, where $n_p'$ is the number of
saddle node bifurcations that correspond to pitchfork bifurcations
for the linear polarization case. The different filling patterns
correspond to values of $n_p$ or $n_p'$ equal to 0 (diagonal lines),
1 (horizontal lines) and 2 (white).} \label{barlam}
\end{figure}

An analysis of Eq.\ (\ref{IntensETA}) indicates that the symmetric
solution ($ I_+ = I_-$) always exists, but there may be more
solutions. We found, in terms of  bifurcations of
intensity, a total of 8 possible cases. These 8 possible cases
define, in the $\alpha \Theta$-$\bar{B}$ plane, 8 regions, that are
limited by 4 curves, as shown in Figure \ref{barlam}(a).

The symmetric solution (regardless if asymmetric solutions exist or
not) presents bistability  if $ \alpha \Theta> \sqrt{3} $,
independently of $ \bar{B} $. The bistability region (i.e. the range
of values of $I_{in}$ for which there are two stable solutions of
$I_\pm$ ) is limited by a pair of saddle-node bifurcations. It has
been extensively studied by other authors
(\cite{kockaert,kockaert2,hoyuelos,moloney,nosotros,lugiato}).

\begin{figure}
\includegraphics[width=13cm]{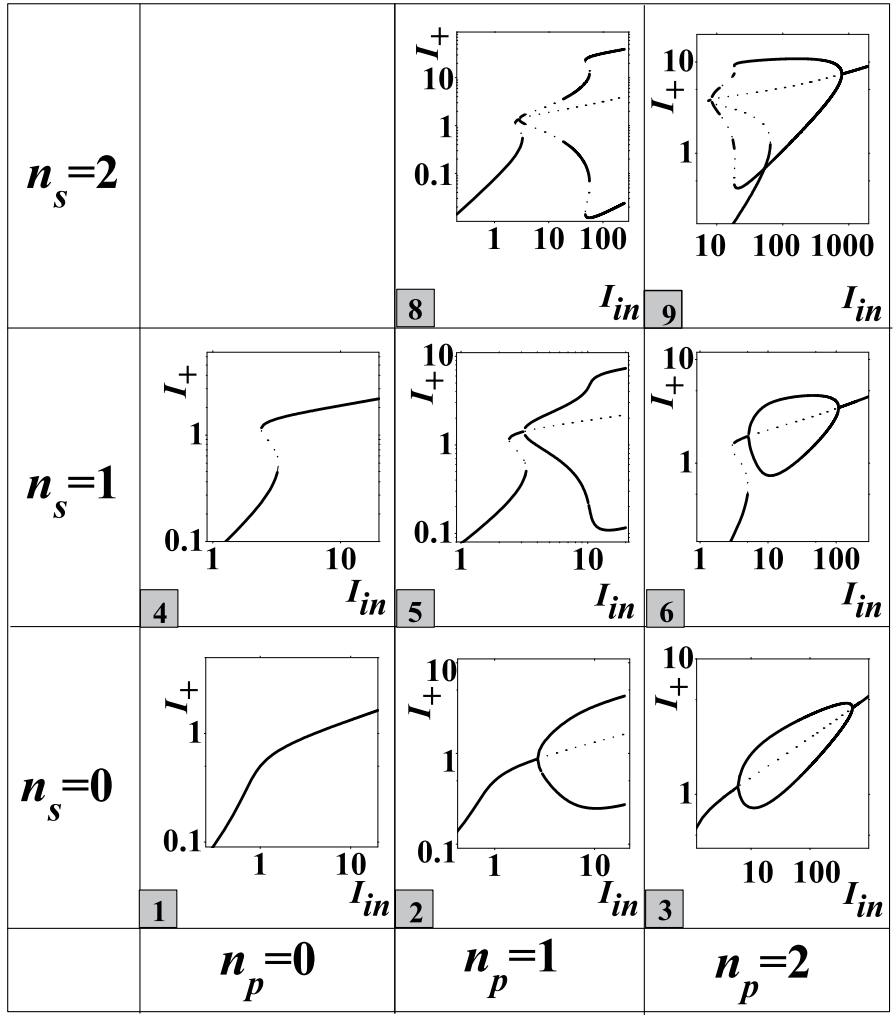}
\caption{Solutions for $\lambda=1/2$.  Full curves were used for
showing stable solutions, and dotted curves for unstable solutions. A
number $N=1+ n_p + 3 n_s$ was assigned to each solution (see caption
of Fig.\ \ref{barlam} for details). Used parameters $(\bar{B}, \alpha
\Theta)$ are: $N=1$, $(0.4,1)$; $N=2$, $(1.2,1)$; $N=3$, $(0.9,1)$; $N=4$,
$(0.4,2.5)$; $N=5$,
$(1.2,2.5)$; $N=6$, $(0.6,3)$; $N=8$, $(1.8,2.5)$;
$N=9$, $(0.5,7.5)$.}\label{sols}
\end{figure}

There might be asymmetric solutions $I_+ > I_-$, and analogous
solutions with $I_+ < I_-$. From Eq.\ (\ref{IntensETA}), looking for
an expression for $ I_-$, which is real and positive, depending on
$I_ +$ for the case $ I_- \neq I_ + $, it is found that: for
$\bar{B}>1$, there is a pitchfork bifurcation in $I_{in}^*$ where two
asymmetric solutions emerge and exist for all $I_{in} > I_{in}^*$
(solutions 2, 5 and 8 in Fig.\ \ref{sols}); for $\bar{B}<1$ and
$\alpha \Theta
> 2 \sqrt{1-\bar{B}}/ \bar{B} $, asymmetric solutions appear between
two values of $ I_{in} $, in which pitchfork bifurcations occur
(solutions 3, 6 and 9 in Fig.\ \ref{sols}); and for $\bar{B}<1$ and
$\alpha \Theta < 2 \sqrt{1-\bar{B}}/ \bar{B} $, only symmetric
solutions exist (solutions 1 and 4 in Fig.\ \ref{sols}).

Finally, it may happen that the asymmetric solution presents
bistability, i.e., the asymmetric branches (generated by a pitchfork
bifurcation) have a curling due to a couple of saddle node
bifurcations. To find them, it is necessary to find the values of
$I_+$ on the asymmetric branches where $ \partial I_{in} (I_+, I_-(I_
+))/\partial I_+ = 0 $. For example, in Fig.\ \ref{sols}, solution 8,
there are 4 values of $I_+$ where this condition is met, and in
solution 5 there are no values that verify the condition. The limit
curve in phase space $\alpha\Theta$-$\bar{B}$ can be analytically
obtained.  It is found that these bifurcations arise for values of
$\alpha \Theta$ greater than $2 \sqrt{3}/ \bar{B}$.

It is not possible to have a symmetric solution with two pairs of
saddle-node bifurcations. For this to happen, we should have 4
points where $\partial I_{in} (I_+, I_-= I_+) / \partial I_+ = 0 $.
This can not be achieved because $ I_{in} $ is a polynomial of
degree 3 in $ I_+ $ when $ I_-= I_+ $, and this is the reason why
there is an empty square in Fig. \ref{sols}.

The possible cases can be organized in a simple way: for a given set
of parameters, we define $ n_p $ as the number of values of $ I_{in}
$ where there are pitchfork bifurcations and $ n_s$ as the number of
regions where bistability  arises due to a couple of saddle-node
bifurcations. There may be only the symmetric solution ($ n_p =  0$),
symmetric solution and asymmetric solution between two values of
incident field ($ n_p =  2$) or asymmetric solution starting from an
incident field value ($ n_p = 1 $).  Independently of the number of
pitchfork bifurcations $n_p$, there may be $n_s = 0$, 1 or 2.

Examples of typical solutions are shown in Fig.\ \ref{sols}, which
also indicates whether the various branches are stable or not under
homogeneous perturbations. In order to classify the solutions with a
unique index we  assign a number, given by $ N = 1 + n_p +3 n_s $, to
each possible case.

\subsection{Circular Polarization,  $\lambda=1$}

From Eq.\ (\ref{IntensETA}),  it can be shown that if $\lambda =1$,
$I_-$ is equal to $0$. For $\alpha \Theta < \sqrt{3}$ there is only
one possible solution, and, for $ \alpha \Theta \geq \sqrt{3}$ there
is the well known bistability generated by a couple of saddle-node
bifurcations. Possible solution regions are shown in Fig.\
\ref{barlam}(d).  These two only possible solutions for circular
polarization are equivalent to solutions 1 and 4 for linear
polarization (see Fig.\ \ref{sols}).

\subsection{Elliptical Polarization, $1/2<\lambda<1$}

As we have seen in the previous subsections, there are many possible
kinds of solutions for $\lambda=1/2$, but only two for $\lambda=1$.
For $1/2<\lambda<1$, up to  11 kinds of solutions were found, but the
boundaries of them are related to the roots of degree 9 polynomials,
 difficulting the derivation of analytical results. However, a
numerical search was made for different values of $ \lambda $ and
some results are shown in Fig.\ \ref{barlam}(b) (for $\lambda=0.6 $),
and (c) (for $\lambda=0.8 $). An analogy between the types of
solution for $ \lambda = 0.5 $ and solution types for $ 1 / 2 <
\lambda <1 $ can be drawn, having in mind some considerations. First,
there is no longer a symmetric solution: there is only one solution
---the \emph{connected} solution--- that exists for $I_{in} \rightarrow 0
$ and is continuous for all values of $I_{in}$. Furthermore, instead
of appearing  or disappearing a couple of asymmetric solutions (as
happened for $\lambda=1/2$ through a pitchfork bifurcation), now a
couple of unconnected solutions appear or disappear through
saddle-node bifurcations.  For $I_{in} \rightarrow 0 $, if
$\lambda=1/2$, only the symmetric solution exists, if $\lambda>1/2$,
only the connected solution exists.

All posible cases can be organized in a simple way similar to what we
did for $ \lambda = 1/2 $. We define $ n_p'$ as the number of values
of $ I_{in} $ where there is a saddle-node bifurcation that makes a
couple of unconnected solutions appear or disappear (these
bifurcations tend to pitchfork bifurcations in the limit $\lambda
\rightarrow 1/2$) and $ n_s $ as the number of regions where
bistability occurs due to a couple saddle-node bifurcations.

The possibilities are: only one connected solution ($ n_p'= 0 $); the
connected solution and a couple of unconnected  solutions between two
values of incident field, produced by two  saddle-node bifurcations
($ n_p'=  2$); and the connected solution plus a couple of
unconnected solutions starting from an incident field value, produced
by a single saddle-node bifurcation ($ n_p'= 1 $). Independently of
the value of $n_p'$, there may be 0, 1 or 2 regions where bistability
emerges due to a couple of saddle-node bifurcations on the connected
branch ($ n_s = 0 $, $1 $ or $ 2$, respectively), or there could be
two pairs of  saddle-node bifurcations on the connected branch and
one on an unconnected branch ($n_s =  3$). In order to enumerate the
cases with a single index, we define $ N = 1 + n_p'+3 n_s $. No cases
have been found with $ n_s =  3$ and $ n_p'= 0 $.

\begin{figure}
\includegraphics[width=13cm]{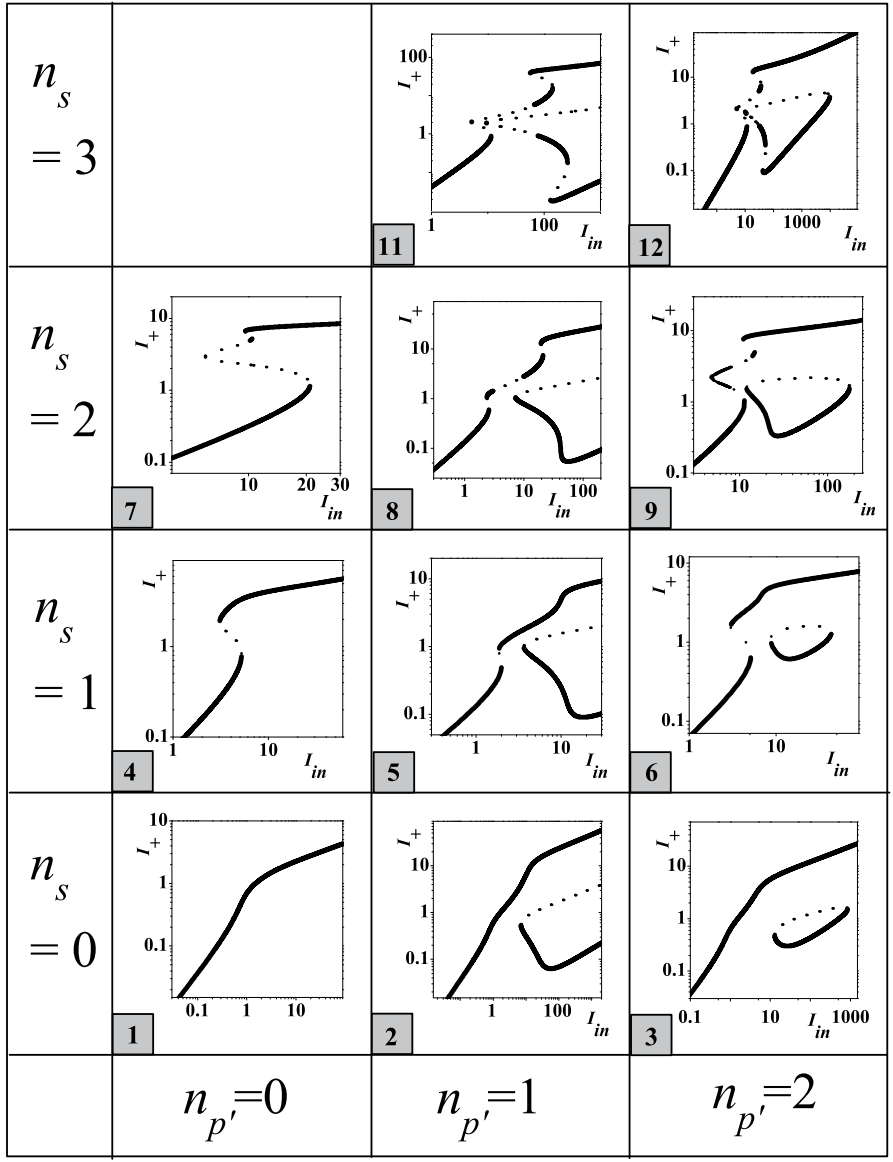}
\caption{Examples of different solutions for $1/2<\lambda<1$. Full
(dotted) line was used for showing stable (unstable) solutions. The
number $N=1+ n_{p}' + 3 n_s   $ was assigned to each solution, where
$n_s$ is the amount of couples of saddle-node bifurcations, and
$n_{p}'$ is the amount of saddle-node bifurcations where an
unconnected couple of solutions start or end. Used parameters
$(\bar{B}, \alpha \Theta, \lambda)$ were: $N=1$, $(0.4,1,0.7)$;
$N=2$, $(1.8,1,0.7)$; $N=3$, $(1.5,1,0.7)$; $N=4$, $(0.4,3,0.7)$;
$N=5$, $(1.4,2,0.55)$; $N=6$, $(0.8,3,0.6)$; $N=7$, $(0.6,5,0.7)$;
$N=8$, $(1.7,2.2,0.7)$; $N=9$, $(1,4,0.7)$; $N=11$, $(1.8,4,0.7)$;
$N=12$, $(1.4,4,0.7)$.} \label{solslam}
\end{figure}

There is one case in which there is some ambiguity in the above
classification: $n_p'= 1$ (unconnected branches) and $n_s = 2$ (two
regions of bistability).  This case corresponds to two qualitatively
different solutions: one in which the two bistability regions are on
the connected solution (this is the case shown in Fig.\
\ref{solslam}, solution 8) and the other in which one bistability
region is on the connected solution and the other solution is on an
unconnected branch. We did not find an equivalent case with $ n_ p'=
2$ and $ n_s = 2 $ where one of the bistability regions is on the
lower branch of
 solutions.

Examples of typical solutions are presented in Figure \ref{solslam}.

There is a last relevant point about  solutions with $\lambda > 1/2$:
if $ n_p' \ge 1 $  a stable  unconnected solution with  $I_+ <I_-$
may occur. This is a counterintuitive result, since the pumping in
the `$+$' polarization is greater than the  one in the `$-$'
polarization. An example of this phenomenon is analyzed in the next
section.

\section{Numerical Integration Results}
\label{simul}

In this section, the temporal evolution of homogeneous solutions of
Eq.\ (\ref{evol}) is studied. In order to do that, numerical
simulations applying 4th order Runge-Kutta method are used.

We found that unstable regions in   $ I_+ (I_{in}) $ graphs are due
to three possible reasons. Whenever there is a pair of saddle-node
bifurcations, the middle branch  will be unstable, and the other two
are expected to be stable, as can be seen in, for example, Fig.\
\ref{sols} for $N = 4$, or Fig.\ \ref{solslam} for $N = 4$. If there
is a pitchfork bifurcation (or saddle node that generates unconnected
solutions), the symmetric branch (or the unconnected middle branch,
i.e., the branch closest to the connected solution) will be unstable,
and it is expected that the other two solutions are stable, as can be
seen in, for example, Fig.\ \ref{sols} for $ N = 2$ and $ 3 $ or
Fig.\ \ref{solslam} for the same values of $ N $. Finally, there may
be areas where we expect a particular branch to be stable, but the
solution becomes unstable due to a Hopf bifurcation. Examples of
regions where Hopf bifurcations occur can be seen in Fig.\ \ref{sols}
for $N = 8$  and 9, and Fig.\ \ref{solslam} for $N = 8$, 9, 11 and
12.

\begin{figure}
\includegraphics[width=16cm]{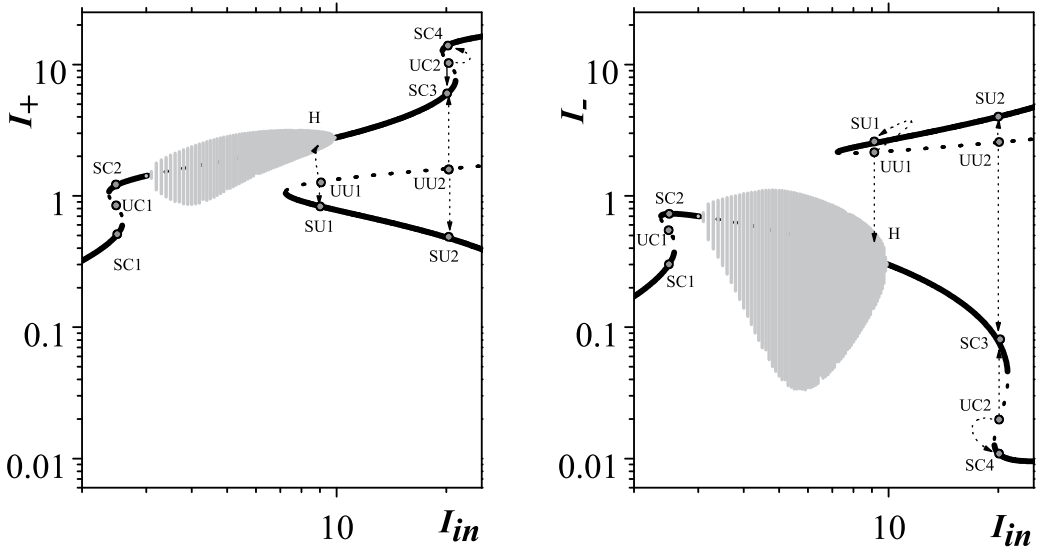}
\caption{Example of evolution of unstable solutions (parameters as in
Fig.\ \ref{solslam}, $N=8$). It can be seen that, for the stable
unconnected solution, $I_- > I_+$ although the input field has
$|A_{in+}|^2 > |A_{in-}|^2$. Full (dotted) line is for stable
(unstable) stationary solutions, and gray regions are for
oscillations due to a Hopf bifurcation. Arrows show the evolution of
a perturbed unstable solution for some typical cases. For $I_{in} =
2.5$, the unstable solution related to a couple of saddle-node
bifurcations
---UC1--- evolves into one of the two stable solutions
---SC1 or SC2---. For $ I_{in} = 9$, if the system starts at the
connected unstable solution ---H---, it evolves towards the Hopf
oscillation; if the system starts at the unconnected unstable
solution ---UU1--- , it can evolve to the lower stable solution
---SU1--- or to the Hopf bifurcation ---H---. For $ I_{in} =20 $,
perturbations around the unstable point of the connected solution
---UC2--- make the system evolve to the stable points of the same
connected solution ---SC3 or SC4---; the unconnected unstable branch
---UU2--- evolves towards the  stable unconnected branch ---SU2--- or
towards the closest stable point of the connected solution ---SC3, it
does not evolve towards SC4---.} \label{ejemplo}
\end{figure}

Numerical results show that, if for a given value of $I_{in}$ there
is only one unstable solution, due to a couple of saddle-node
bifurcations or a  pitchfork bifurcation (or saddle-node that
generates unconnected solutions), small perturbations in the
solution make it evolve to one of the two stable solutions, and the
chosen solution depends only on the disturbance. Also, when there is
only one solution and it is unstable due to a Hopf bifurcation, a
disturbance takes the system to the oscillatory solution. The latter
case occurs, for example, with the solution presented in Fig.
\ref{solslam} for $N= 8$ and $I_{in} $ between 3 and 7.2. Fig.
\ref{ejemplo} shows the same case, indicating with arrows the
possible values of $ I_+ $ and $ I_-$ that the unstable solution
takes after it is perturbed. It is important to note that the Hopf
oscillation occurs in a four-dimensional space generated by the real
and imaginary parts of $ A_ + $ and $ A_-$, and the present graph shows
only the square modules of those fields.

Interesting behaviors occur when, for a given $ I_{in} $ value, there
are two or more unstable solutions, due to bifurcations generated by
instabilities of the same or different types. For instance, it can be
seen in Fig. \ref{ejemplo}, for $I_{in}=20$, that the unstable
unconnected solution evolves towards the stable unconnected solution
or to the closest stable connected solution (but it does not evolve
to the other stable connected solution). Also, in the same figure and
for the same value of $I_{in}$, the unstable connected solution (the
dotted part of the connected solution, which occurs due to a couple
of saddle node bifurcations on the connected solution) evolves
towards one of the two stable connected solutions (it does not go
towards the unconnected stable solution).

\section{Conclusions}
\label{concl}

We described the evolution of the electromagnetic field in a  ring
cavity with flat mirrors, filled with a Kerr-type electric and
magnetic nonlinear material with positive or negative refraction
index. We took into account the degree of freedom of the polarization
of the incident field. Starting from two pairs of coupled nonlinear
Schr\"{o}dinger equations (for the circularly left and right polarized
components of the electric and magnetic fields), we found a
proportionality relation between electric and magnetic field
components, so that the description was reduced to just two Lugiato
Lefever equations.

Considering that the effective magnetic response can be nonlinear,
and taking into account the polarization degree of freedom of the
incident beam, a rich variety of possible bifurcations in homogeneous
stationary solutions was found, depending on the values of the
constants that describe the system. These solutions were classified and
exemplified. The number of qualitatively different solutions that
were obtained is large, and our classification in terms of the number
of possible bifurcations simplifies the description.

In the case of linear polarization ($ \lambda =  1 / 2$), we found
that the solutions $ I_+ $ and $ I_-$ depending on $ I_{in}  $, could
present between 0 and 2 pitchfork bifurcations and between 0 and 2
pairs of saddle-node bifurcations, yielding 8 possible cases. We
determined the boundaries between regions of each type of solution in
the parameter space given by $ \alpha \Theta $ and $ \bar{B} $. The
same analysis was performed for circular polarization ($ \lambda = 1
$).

For elliptical polarization ($ 1 / 2 < \lambda <1 $), we found 11
possible cases based on the number of bifurcations. We numerically
explored regions of parameters that generated each case.

We also numerically analyzed how unstable homogeneous solutions
evolve after a disturbance, and found regions where  Hopf
instabilities take place.

Finally, for elliptical polarization there are some sets of
parameters where, taking an incident field polarized mostly in the
`$+$' component, it is possible to obtain stable solutions where the
`$-$' polarization component prevails.

In summary, we presented an analysis of the possible homogeneous
stationary solutions of the electromagnetic field in a ring cavity
containing a material with electric and magnetic nonlinearities, and
taking into account the polarization degree of freedom of the light.
Let us note that even for linearly polarized incident light,
elliptically polarized solutions can appear, so that considering only
a scalar equation may cause the removal of a possible solution.
Despite the large number of parameters of the original description, a
classification of the possible solutions for a given polarization
could be obtained taking into account only two parameters: $\alpha$
(related to the sign of the electric and magnetic nonlinear
susceptibilities) times $\Theta$ (the scaled detuning) and $\bar{B}$
(related to the electric and magnetic nonlinear constants $B_E$ and
$B_M$).  Results presented here can be useful when ring cavities are
used in optical switching or information storage.

\begin{acknowledgments}
This work was partially supported by Consejo Nacional de
Investigaciones Cient\'{\i}ficas y T\'{e}cnicas (CONICET, Argentina, PIP 0041
2010-2012).
\end{acknowledgments}

\end{document}